\documentclass[aps,prb,twocolumn,amsmath,amssymb,nofootinbib,superscriptaddress,floatfix]{revtex4-1}
\usepackage{amsmath}
\usepackage{amssymb}
\usepackage{amsthm}
\usepackage{graphicx}%
\usepackage{color}

\usepackage{amsthm}

\theoremstyle{plain}

\newtheorem{obs}{Observation}
\newtheorem{ques}{Question}
\newtheorem{cor}{Corollary}

\theoremstyle{definition}

\newtheorem{defn}{Definition}

\newcommand\ket[1]{\ensuremath{|#1\rangle}}
\newcommand\bra[1]{\ensuremath{\langle#1|}}

\usepackage{xspace}

\makeatletter
\newcommand*{\etc}{%
    \@ifnextchar{.}%
        {etc}%
        {etc.\@\xspace}%
}
\makeatother

\newcommand\Z{\mathbb{Z}}

\newcommand\Ref{Ref}

\newcommand\eqn{\eqref}

\newcommand\Tr{\text{Tr}}

\begin{document}

\title{Gapped quantum liquids and topological order,\\ 
stochastic local transformations and emergence of unitarity}


\author{Bei Zeng}%
\affiliation{Department of Mathematics \& Statistics, University of
  Guelph, Guelph, Ontario, Canada}%
\affiliation{Institute for Quantum Computing, University of Waterloo,
  Waterloo, Ontario, Canada}%
\affiliation{Canadian Institute for Advanced Research, Toronto, Ontario, Canada}%

\author{Xiao-Gang Wen}
\affiliation{Department of Physics, Massachusetts Institute of Technology, Cambridge, Massachusetts 02139, USA}
\affiliation{Perimeter Institute for Theoretical Physics, Waterloo, Ontario, N2L 2Y5 Canada} 


\begin{abstract}
In this work we present some new understanding of topological order, 
including three main aspects: (1) It was believed that
classifying topological orders corresponds to classifying gapped quantum
states. We show that such a statement is not precise.  We introduce the concept
of \emph{gapped quantum liquid} as a special kind of gapped quantum states that
can ``dissolve'' any product states on additional sites.  Topologically ordered
states actually correspond to gapped quantum liquids with stable ground-state
degeneracy.  Symmetry-breaking states for on-site symmetry are also gapped
quantum liquids, but with unstable ground-state degeneracy.  (2) We point out
that the universality classes of generalized local unitary (gLU)
transformations (without any symmetry) contain both topologically ordered
states and symmetry-breaking states. This allows us to use a gLU invariant --
topological entanglement entropy -- to probe the symmetry-breaking properties
hidden in the exact ground state of a finite system, which does not break any
symmetry. This method can probe symmetry- breaking orders even without knowing
the symmetry and the associated order parameters.  (3) The universality classes
of topological orders and symmetry-breaking orders can be distinguished by
\emph{stochastic local (SL) transformations} (i.e.\ \emph{local invertible
transformations}): small SL transformations can convert the symmetry-breaking
classes to the trivial class of product states with finite probability of
success, while the topological-order classes are stable against any small SL
transformations, demonstrating a phenomenon of emergence of unitarity.  This
allows us to give a new definition of long-range entanglement based on SL
transformations, under which only topologically ordered states are long-range
entangled.

\end{abstract}

\maketitle

\section{Introduction}

Topological order was first introduced as a new kind of order beyond Landau
symmetry breaking theory~\cite{Wtop,WNtop,Wrig}. At the beginning, it was
defined by (a) the topology-dependent ground-state degeneracy~\cite{Wtop,WNtop}
and (b) the non-Abelian geometric phases of the degenerate ground
states\cite{Wrig,KW9327}, where both of them are \emph{robust against any local
perturbations} that can break any symmetry~\cite{WNtop}. This is just like
that superfluid order is given by zero-viscosity and quantized vorticity, which
are robust against any local perturbations that preserve the $U(1)$ symmetry.
Chiral spin liquids~\cite{KL8795,WWZ8913}, integral/fractional quantum Hall
states~\cite{KDP8094,TSG8259,L8395}, $\Z_2$ spin
liquids~\cite{RS9173,W9164,MS0181}, non-Abelian fractional quantum Hall
states~\cite{MR9162,W9102,WES8776,RMM0899} \etc, are examples of topologically
ordered phases.

Microscopically, superfluid order is originated from boson or fermion-pair
condensation.  So it is natural for us to ask: what is the  microscopic origin
of topological order?  What is  the  microscopic origin of robustness against
\emph{any} local perturbations?  Recently, it was found that, microscopically,
topological order is related to long-range entanglement~\cite{LW0605,KP0604}. 
In fact, we can regard topological order as patterns of long-range
entanglement~\cite{CGW1038} defined through local unitary (LU)
transformations~\cite{LWstrnet,VCL0501,V0705}.

In this paper, we will discuss in more detail the relation between topological
order and many-body quantum entanglement.  We first point out that the
topologically ordered states are not arbitrary gapped states, but belong to a
special kind of gapped quantum states, called \emph{gapped quantum liquids}. We
will give a definition of gapped quantum liquids.  Haah's cubic model may be an
example of gapped quantum states which is not a gapped quantum
liquid~\cite{haah2011local}.  

The notion of gapped quantum liquids can also be applied to solve the problem
of how to take the thermodynamic limit for systems without translation
symmetry.  In general, in the presence of strong randomness, the thermodynamic
limit is not well defined (without impurity average). We show that for gapped
quantum liquids, the thermodynamic limit is well defined even without impurity
average. Consequently, the notions of quantum phases and quantum phase
transitions are well defined for gapped quantum liquids.

We then show that the equivalence classes of gLU transformations, not only
describe topologically ordered states, but also include the ground states of
symmetry-breaking phases, where the exact symmetric ground states have
entanglement of the Greenberger-Horne-Zeilinger~\cite{greenberger1989going}
(GHZ) form.  This allows us to use a gLU invariant -- topological entanglement
entropy -- to probe the symmetry-breaking properties hidden in the exact ground
state of a finite system, which is invariant under the symmetry transformation.
Note that, to use the topological entanglement entropy to probe the symmetry
breaking states, \emph{we do not need to know the symmetry or the
symmetry-breaking order parameters}.  Usually, one needs to identify the
symmetry-breaking order parameters and compute their long-range correlation
functions to probe the symmetry-breaking properties hidden in the symmetric
exact ground-state wavefunction.

We further show that many-body states with GHZ-form entanglement are
convertible to product states with a finite probability under stochastic local
(SL) transformations, which are local invertible transformations that are not
necessarily unitary. In contrast, topologically ordered states are not
convertible to product states via small SL transformations.  This allows us to
give a new definition of long-range entanglement based on SL convertibility to
product states, under which only topologically ordered states have long-range
entanglement.  Moreover, we show that the topological entanglement entropy for
topological orders is stable under small SL transformations but unstable for
symmetry-breaking orders.

For topologically ordered states, the robustness of the ground-state degeneracy
and the robustness of the unitary non-Abelian geometric phases against any
(small) SL transformations (i.e.\ local non-unitary transformations) reveal the
phenomenon of \emph{emergence of unitarity}: even when the bare quantum
evolution at lattice scale is non-unitary, the induced adiabatic evolution in
the ground-state subspace is still unitary. In this sense, topological order
can be defined as states with emergent unitary for non-unitary quantum
evolutions.  The phenomenon of emergence of unitarity may have deep impact in
the foundation of quantum theory, and in the elementary particle theory, since
the emergence/unification of gauge interaction and Fermi statistics is closely
related to topological order and long-range entanglement~\cite{LW0316}. (The
emergence of unitarity is also discussed in the $N=4$ supersymmetric Yang-Mills
scattering amplitudes in the planar limit~\cite{AT1307}.)

\section{Gapped quantum liquids and topological order}

\subsection{Gapped quantum system and gapped quantum phase}

Topologically ordered states are gapped quantum states. 
To clarify the concept of gapped quantum
states, we first define gapped quantum system.  Since a gapped system may
have gapless excitations on the boundary (such as quantum Hall systems), so to
define gapped Hamiltonians, we need to put the Hamiltonian on a space with no
boundary.  Also, system with certain sizes may contain non-trivial excitations
(such as a spin liquid state of spin-1/2 spins on a lattice with an ODD number
of sites), so we have to specify that the system has a certain sequence of
sizes when we take the thermodynamic limit.
\begin{defn} \textbf{Gapped quantum system}\\
Consider a local Hamiltonian of a qubit system on graphs with no boundary, with
finite spatial dimension $D$.  If there is a sequence of sizes of the system
$N_k$, $N_k\to \infty$, as $k\to \infty$, such that the size-$N_k$ system has
the following ``gap property'', then the system, defined by the Hamiltonian sequence $\{H_{N_k}\}$, is said to be gapped.  Here
$N_k$ can be viewed as the number of qubits in the system.
\end{defn} \noindent
\begin{defn} \textbf{Gap property}\\
There is a fixed $\Delta$ (i.e.\ independent of $N_k$) such that\\
 (1) the size-$N_k$ Hamiltonian has no eigenvalue in an energy window of size $\Delta$;\\
 (2) the number of eigenstates below the energy window does not depend on 
     $N_k$;\\ 
 (3) the energy splitting of those eigenstates below the energy window
approaches zero as $N_k \to \infty$.  
\end{defn} \noindent 
Note that the notion of ``gapped quantum system'' cannot be even defined for a
single Hamiltonian. It is a property of a sequence of Hamiltonians,
$\{H_{N_k}\}$, in the large size limit $N_k \to \infty$.  In this paper, the
term ``a gapped quantum system'' refers to a sequence of Hamiltonians,
$\{H_{N_k}\}$, that satisfies the above properties.

Now we can give a precise definition for
\begin{defn}
\textbf{Ground-state degeneracy and ground-state subspace}\\ 
The number of eigenstates below the energy window becomes the ground-state
degeneracy of the gapped system.  (This is how the ground-state degeneracy of a
topologically ordered state is
defined~\cite{Wtop,WNtop,Wrig,bravyi2010topological}.) The states below the
energy window span the \emph{ground-state subspace}, which is denoted as
$\mathcal{V}_{N_k}$.
\end{defn} \noindent

\begin{figure}[tb]
\begin{center}
\includegraphics[scale=0.5]{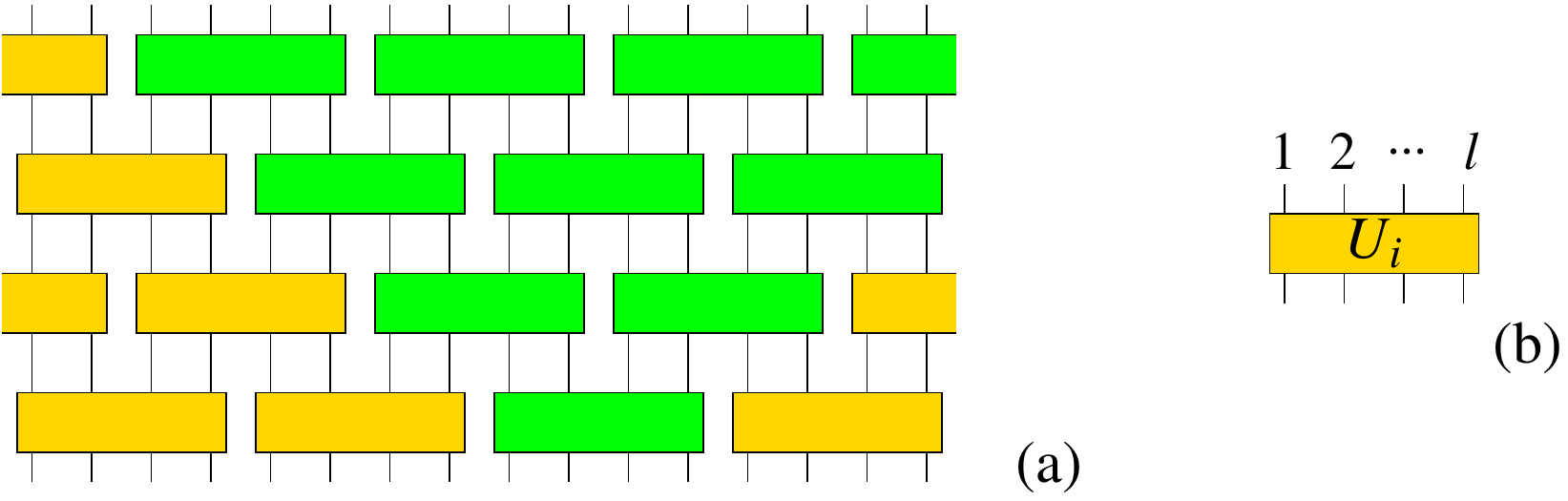}
\end{center}
\caption{
(Color online)
(a) A graphic representation of a quantum circuit, which is form by (b) unitary
operations on blocks of finite size $l$. The green shading represents a causal
structure.
}
\label{qc}
\end{figure}

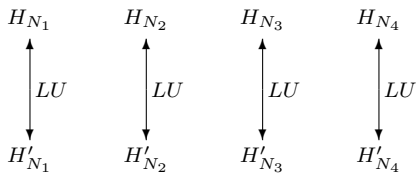
\begin{figure}[h!]
\vskip-\baselineskip
\centerline{\footnotesize\unitlength 0.8\unitlength%
\begin{tabular}{cccc}
\begin{picture}(50,80)(0,-40)
\put(25,33){\makebox(0,0){$H_{N_1}$}}
\put(35,0){\makebox(0,0){$LU$}}
\put(25,0){\vector(0,1){24}}
\put(25,0){\vector(0,-1){24}}
\put(25,-33){\makebox(0,0){$H'_{N_1}$}}
\end{picture}
&
\begin{picture}(50,80)(0,-40)
\put(25,33){\makebox(0,0){$H_{N_2}$}}
\put(35,0){\makebox(0,0){$LU$}}
\put(25,0){\vector(0,1){24}}
\put(25,0){\vector(0,-1){24}}
\put(25,-33){\makebox(0,0){$H'_{N_2}$}}
\end{picture}
&
\begin{picture}(50,80)(0,-40)
\put(25,33){\makebox(0,0){$H_{N_3}$}}
\put(35,0){\makebox(0,0){$LU$}}
\put(25,0){\vector(0,1){24}}
\put(25,0){\vector(0,-1){24}}
\put(25,-33){\makebox(0,0){$H'_{N_3}$}}
\end{picture}
&
\begin{picture}(50,80)(0,-40)
\put(25,33){\makebox(0,0){$H_{N_4}$}}
\put(35,0){\makebox(0,0){$LU$}}
\put(25,0){\vector(0,1){24}}
\put(25,0){\vector(0,-1){24}}
\put(25,-33){\makebox(0,0){$H'_{N_4}$}}
\end{picture}
\end{tabular}}
\caption{
The two rows of Hamiltonians describe two gapped quantum systems.  The two rows
connected by LU transformations represent the equivalence relation between the
two gapped quantum systems, whose equivalence classes are gapped quantum
phases. We may view $H_N$ as a projector that defines its ground-state subspace.
The ground-state subspaces of two  equivalent systems are connected by
the LU transformations.
}
\label{gapLUH}
\end{figure}

Now, we would like to define gapped quantum phase. First, we introduce
\begin{defn}
\textbf{Local unitary (LU) transformation}\cite{CGW1038}\\ 
An LU transformation can be given by a quantum circuit as shown 
in Fig.~\ref{qc}. 
An LU transformation is given by a finite layers (i.e.\
the number of layers is 
a constant that is independent of the system size) of piecewise
local unitary transformations
\begin{equation*}
 U^M_{circ}= U_{pwl}^{(1)} U_{pwl}^{(2)} \cdots U_{pwl}^{(M)}
\end{equation*}
where each layer has a form
\begin{equation*}
U_{pwl}= \prod_{i} U^i.
\end{equation*}
Here $\{ U^i \}$ is a set of unitary operators that act on non-overlapping
regions. The size of each region is less than a finite number $l$.
\end{defn} \noindent
Two gapped systems connected by a LU transformation can deform into each other smoothly without closing the energy gap, and thus belong to the same phase. This leads us to define
\begin{defn}
\textbf{Gapped quantum phase}\\
\label{gPhase}
Two gapped quantum systems $\{H_{N_k}\}$ and $\{H'_{N_k}\}$ are equivalent if
the \emph{ground-state subspaces} of $H_{N_k}$ and $H'_{N_k}$ are connected by
LU transformations for all $N_k$. The equivalence classes of the above
equivalence relation are the gapped quantum phases (see Fig. \ref{gapLUH}).
\end{defn}

It is highly desired to identify topological orders as gapped quantum phases,
since both concepts do not involve symmetry.  In the following, we will show
that gapped quantum phases, sometimes, are not well behaved in the
thermodynamic limit.  As a result, it is not proper to define topological
orders as gapped quantum phases.  To fix this problem, we will introduce the
concept of gapped quantum liquid phase.

\subsection{Gapped quantum liquid system and gapped quantum liquid phase}

Why gapped quantum systems may not be well behaved in the thermodynamic
limit? This is because the Hamiltonians with different sizes are not related
(see Fig. \ref{gapLUH}) in our definition of gapped quantum systems.  As a
result, we are allowed to choose totally different $H_{N_k}$ and $H_{N_{k+1}}$
as long the Hamiltonians have the same ground-state degeneracy.  For example,
one can be topologically ordered and the other can be symmetry-breaking.  To
fix this problem, we choose a subclass of  gapped quantum systems which is
well-behaved in the thermodynamic limit.  Those gapped quantum systems are
``shapeless'' and can ``dissolve'' any product states on
additional sites to increase its size.
Such gapped quantum systems are called \textit{gapped quantum liquid systems}. 

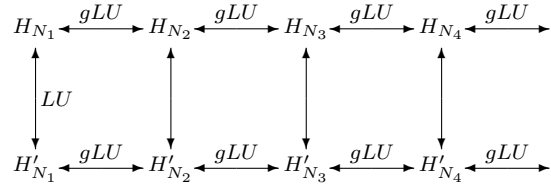
\begin{figure}[hbtp]
\begin{center}
\vskip-\baselineskip
\centerline{\footnotesize\unitlength 0.8\unitlength%
\begin{tabular}{ccccccccc}
\begin{picture}(50,80)(20,-30)
\put(25,33){\makebox(0,0){$H_{N_1}$}}
\put(35,0){\makebox(0,0){$LU$}}
\put(25,0){\vector(0,1){24}}
\put(25,0){\vector(0,-1){24}}
\put(25,-33){\makebox(0,0){$H'_{N_1}$}}
\end{picture}
&
\begin{picture}(4,80)(44,-30)
\put(25,40){\makebox(0,0){$gLU$}}
\put(25,33){\vector(1,0){20}}
\put(25,33){\vector(-1,0){20}}
\put(25,-26){\makebox(0,0){$gLU$}}
\put(25,-33){\vector(1,0){20}}
\put(25,-33){\vector(-1,0){20}}
\end{picture}
&
\begin{picture}(50,80)(20,-30)
\put(25,33){\makebox(0,0){$H_{N_2}$}}
\put(25,0){\vector(0,1){24}}
\put(25,0){\vector(0,-1){24}}
\put(25,-33){\makebox(0,0){$H'_{N_2}$}}
\end{picture}
&
\begin{picture}(4,80)(44,-30)
\put(25,40){\makebox(0,0){$gLU$}}
\put(25,33){\vector(1,0){20}}
\put(25,33){\vector(-1,0){20}}
\put(25,-26){\makebox(0,0){$gLU$}}
\put(25,-33){\vector(1,0){20}}
\put(25,-33){\vector(-1,0){20}}
\end{picture}
&
\begin{picture}(50,80)(20,-30)
\put(25,33){\makebox(0,0){$H_{N_3}$}}
\put(25,0){\vector(0,1){24}}
\put(25,0){\vector(0,-1){24}}
\put(25,-33){\makebox(0,0){$H'_{N_3}$}}
\end{picture}
&
\begin{picture}(4,80)(44,-30)
\put(25,40){\makebox(0,0){$gLU$}}
\put(25,33){\vector(1,0){20}}
\put(25,33){\vector(-1,0){20}}
\put(25,-26){\makebox(0,0){$gLU$}}
\put(25,-33){\vector(1,0){20}}
\put(25,-33){\vector(-1,0){20}}
\end{picture}
&
\begin{picture}(50,80)(20,-30)
\put(25,33){\makebox(0,0){$H_{N_4}$}}
\put(25,0){\vector(0,1){24}}
\put(25,0){\vector(0,-1){24}}
\put(25,-33){\makebox(0,0){$H'_{N_4}$}}
\end{picture}
&
\begin{picture}(4,80)(44,-30)
\put(25,40){\makebox(0,0){$gLU$}}
\put(25,33){\vector(1,0){20}}
\put(25,33){\vector(-1,0){20}}
\put(25,-26){\makebox(0,0){$gLU$}}
\put(25,-33){\vector(1,0){20}}
\put(25,-33){\vector(-1,0){20}}
\end{picture}
\end{tabular}}
\caption{
The two rows define two gapped quantum liquid systems via gLU transformations.
The two rows connected by LU transformations represent the equivalence relation
between two gapped quantum liquid systems, whose equivalence classes are
gapped quantum liquid phases.
}
\label{liquidLUH}
\end{center}
\end{figure}

\begin{defn}
\textbf{Gapped quantum liquid system}\\ 
A gapped quantum liquid system is
a gapped quantum system, described by the sequence $\{H_{N_k}\}$, 
with two additional properties:\\
(1) $0<c_1<(N_{k+1}-N_{k})/N_k < c_2 $ where $c_1$ and $c_2$ are constants that do not depend on the system size. \\
(2) the ground-state subspaces of $H_{N_k}$ and $H_{N_{k+1}}$ are connected by a generalized local unitary
(gLU) transformation (see Fig. \ref{liquidLUH}).  
\end{defn} \noindent

\begin{figure}[h!] 
\begin{center} 
\includegraphics[scale=0.5]{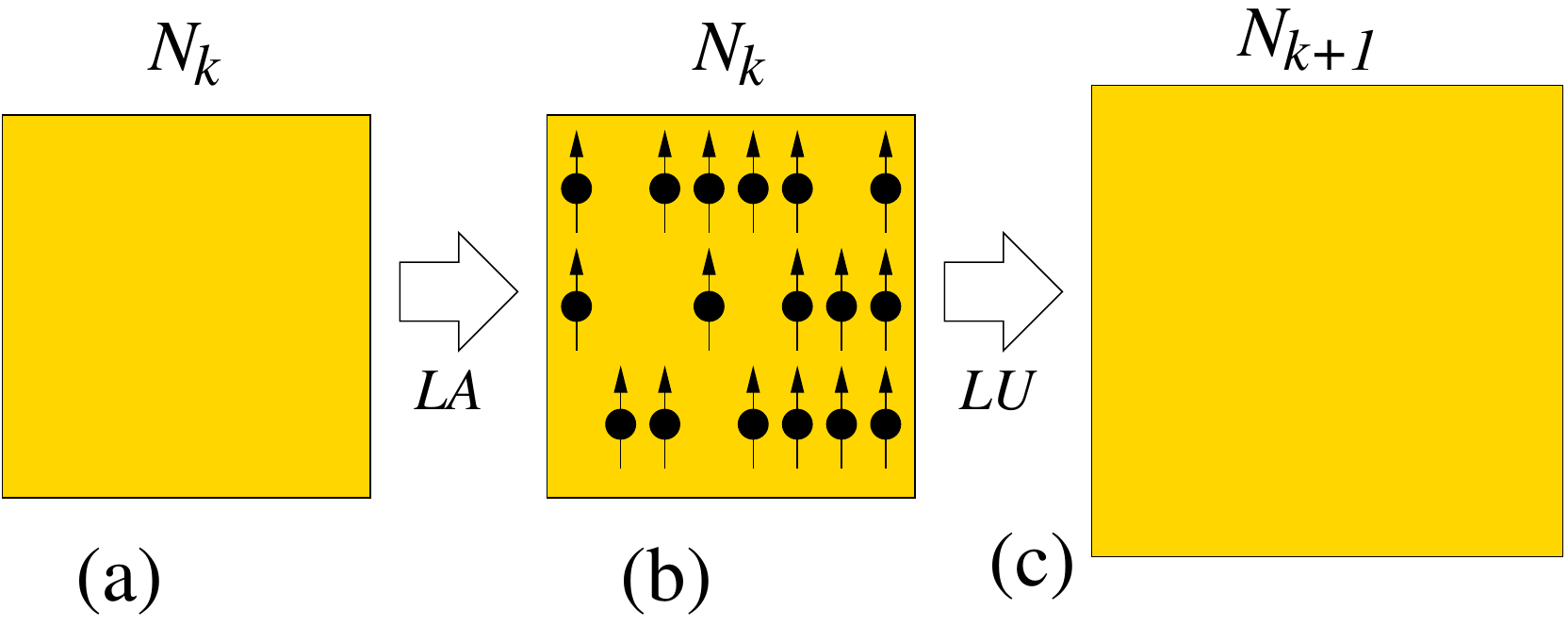} \end{center}
\caption{
(Color online) Two systems (a) and (c), with size $N_k$ and $N_{k+1}$, are
described by $H_{N_k}$ and $H_{N_{k+1}}$ respectively.  (a) $\to$ (b) is an LA
transformation where we add $N_{k+1}-N_k$ qubits to the system $H_{N_k}$ to
obtain the Hamiltonian $H_{N_k} + \sum_i Z_i$ for the combined system (b).
Under the LA transformation, the ground states of $H_{N_k}$ is tensored with a
product state to obtain the ground states of $H_{N_k}+ \sum_i Z_i$.  In (b)
$\to$ (c), we transform the ground-state subspace of $H_{N_k} + \sum_i Z_i$ to
the ground-state subspace of $H_{N_{k+1}}$ via an LU transformation.
} 
\label{NkNk} 
\end{figure}

Fig. \ref{NkNk} explains how we transform $H_{N_k}$ to $H_{N_{k+1}}$ via a
gLU transformation. For the system $H_{N_k}$, we first need to add  
$N_{k+1}-N_k$ qubits. We would like to do this addition ``locally".
That is, the distribution of the added
qubits may not be uniform in space but maintains a finite density
(number of qubits per unit volume). We then define
how to write Hamiltonians after adding particles to the system.

\begin{defn}
\textbf{Local addition (LA) transformation}\\ 
For adding $N_{k+1}-N_k$ qubits to the system $H_{N_k}$ locally,
we consider the Hamiltonian $H_{N_k} + \sum_{i=1}^{N_{k+1}-N_k} Z_i$ 
for the combined system (see Fig. \ref{NkNk}b), where $Z_i$ 
is the Pauli $Z$ operator acting on the
$i^\text{th}$ qubit.  This defines an LA transformation from
$H_{N_k}$ to $H_{N_k} + \sum_{i=1}^{N_{k+1}-N_k} Z_i$. 
\end{defn} \noindent
\begin{defn}
\textbf{gLU transformation}\\ 
If for any LA transformation from $H_{N_k}$ to $H_{N_k} +
\sum_{i=1}^{N_{k+1}-N_k} Z_i$, the ground-state subspace of $H_{N_k} +
\sum_{i=1}^{N_{k+1}-N_k} Z_i$ can be transformed into  the ground-state
subspace of $H_{N_{k+1}}$ via an LU transformation, then we say $H_{N_k}$ and
$H_{N_{k+1}}$ are connected by a gLU transformation.  
\end{defn} \noindent
 
\begin{figure}[tb] 
\begin{center} 
\includegraphics[scale=0.2]{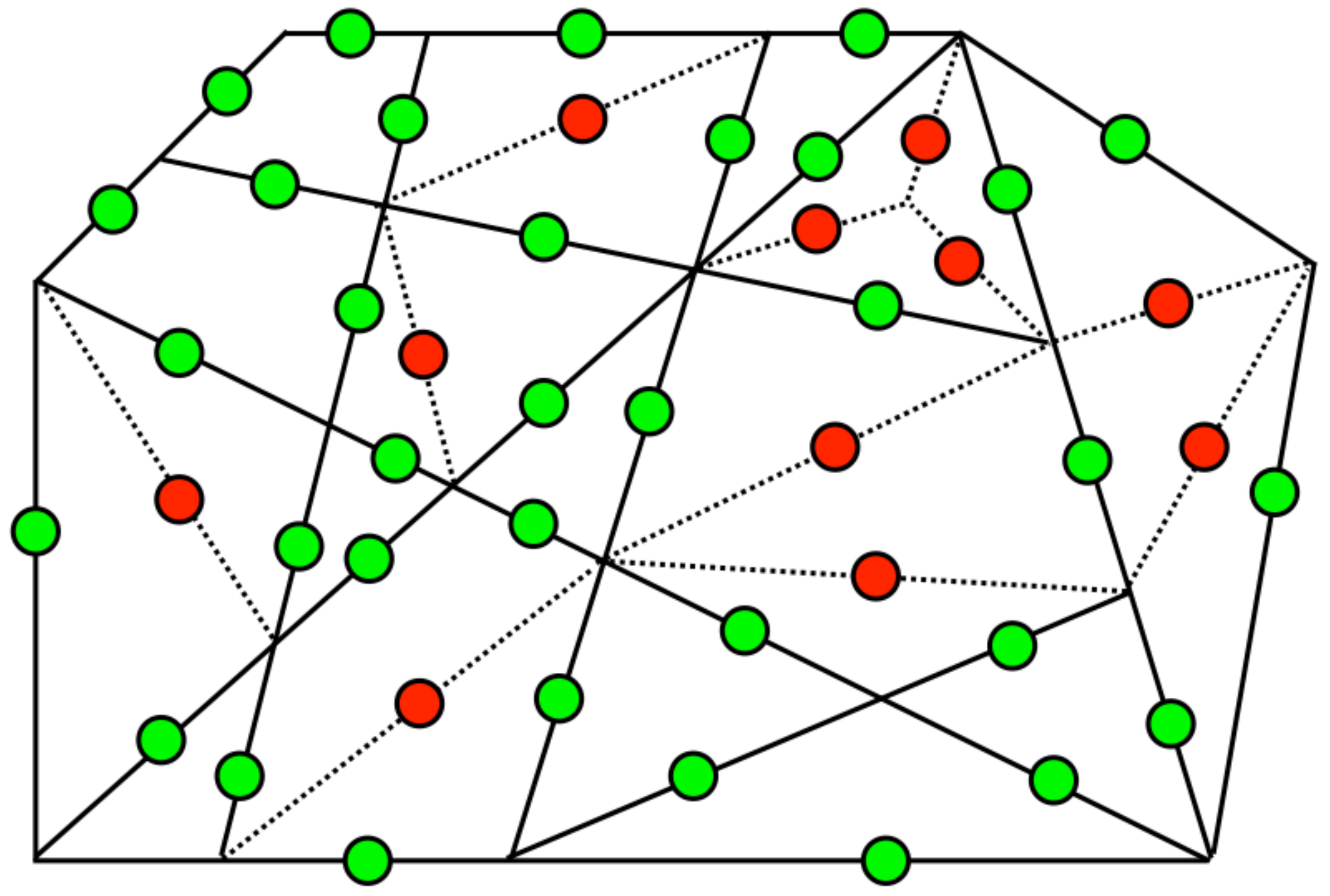} \end{center}
\caption{
Toric code as a gapped quantum liquid: toric code of $N_k$ qubits on an arbitrary 
2D lattice, where the green dots represent qubits sitting on the link of the lattice (given by 
solid lines). By
adding $N_{k+1}-N_{k}$ qubits (red dots), the gLU transformation 
$H_{N_k}\to H_{N_{k+1}}$ `dissolves' the red qubits in the new lattice (with both
the solid lines and dashed lines).
} 
\label{Toric1} 
\end{figure}

According to our definition, the sequence of following Hamiltonians
\begin{equation}
H^\text{trivial-liquid}_{N_k}=-\sum_{i=1}^{N_k}Z_i,
\end{equation}
gives rise to a gapped quantum liquid system. 
The topologically-ordered toric code Hamiltonian $H^\text{toric}_{N_k}$ is also
a gapped quantum liquid, as illustrated in Fig.~\ref{Toric1}. This reveals one
important feature of a gapped quantum liquid -- the corresponding
lattice in general do not have a `shape'
(i.e.\ the system can be defined on an arbitrary lattice with a meaningful thermodynamic limit).

To have an example of a gapped system that is not 
a quantum liquid, 
consider another sequence of Hamiltonians
\begin{equation}
\label{Hnonliquid}
H^\text{non-liquid}_{N_k}=-\sum_{i=1}^{N_k-1}Z_i.
\end{equation}
It describes a gapped quantum system with two-fold degenerate ground states
(coming from the $N_k^\text{th}$ qubit which carries no energy).  However, such
a gapped quantum system is not a gapped quantum liquid system. Because
the labelling of the $N_{k+1}$ qubit is essentially arbitrary,
for some LA transformations, the map from 
$H_{N_k} + \sum_{i=1}^{N_{k+1}-N_k} Z_i$ to $H_{N_{k+1}}$
cannot be local.

Through the above example, we see that a gapped quantum system may not
have a well defined thermodynamic limit (because the low energy property --
the degenerate ground states, is given by an isolated qubit which is not a
thermodynamic property.) Similarly, gapped quantum phase (as defined in
Definition \ref{gPhase}) is not a good concept, since it is not a thermodynamic
property sometimes.  In contrast, gapped quantum liquid system and
gapped quantum liquid phase (defined below in Definition \ref{glPhase}) are a
good concepts, since they are always thermodynamic properties.

We also believe that the cubic code of the Haah model is another example of
gapped quantum system that is not a gapped quantum liquid
system~\cite{haah2011local}.  There exists a sequence of the linear sizes of
the cube: $L_k\rightarrow\infty$, where the ground-state degeneracy is two,
provided that $L_k=2^k-1$ (or $L_k=2^{2k+1}-1$) for any integer $k$, and
correspondingly $N_k=L_k^3$.  However, we do not think that
$H^\text{Haah}_{N_k}$ and
$H^\text{Haah}_{N_{k+1}}$ are connected by an gLU transformation.  Here
$H^\text{Haah}_{N_k}$ is the Hamiltonian of the cubit code of size $N_k$. 

We can now define:
\begin{defn}
\textbf{Gapped quantum liquid phase}\\
\label{glPhase}
Two gapped quantum liquid systems $\{H_{N_k}\}$ and $\{H'_{N_k}\}$ are
equivalent if the ground-state subspaces of
$H_{N_k}$ and $H'_{N_k}$ are connected by LU transformations
for all $N_k$. The equivalence classes of the above equivalence relation
are the gapped quantum liquid phases (see Fig. \ref{liquidLUH}).
\end{defn}

\subsection{Topological order}

Using the notion of gapped quantum liquid phase, we can have a definition of
topological order.  First, we introduce
\begin{defn} \textbf{Stable gapped quantum system}\\
If the ground-state degeneracy of an gapped quantum system
is stable against any local perturbation (in the large $N_k$ limit),
then the gapped quantum system is stable.
\end{defn} \noindent

An intimately related fact to this definition
is that the ground-state subspace of a stable gapped quantum system 
(in the large $N_k$ limit) is a quantum
error-correcting code with macroscopic distance~\cite{bravyi2010topological}. 
This is to say, for
any orthonormal basis $\{\ket{\Phi_i}\}$ of the ground-state subspace, for any local operator
$M$, we have 
\begin{equation}
\label{eq:qcode}
\bra{\Phi_i}M\ket{\Phi_j}=C_M\delta_{ij},
\end{equation}
where $C_M$ is a constant which only depends on $M$~\cite{furukawa2006systematic,furukawa2007reduced,nussinov2009symmetry,
schuch2010peps}.

Note that a gapped quantum liquid system may not be a stable gapped quantum
system. A symmetry-breaking system is an example, which is a  gapped quantum
liquid system but not a stable gapped quantum system (the ground-state
degeneracy can be lifted by symmetry-breaking perturbations).  Also a stable
gapped quantum system may not be a gapped quantum liquid system.  A non-Abelian
quantum Hall states\cite{MR9162,W9102} with traps\cite{LW1384} that trap
non-Abelian quasiparticles is an example.  Since the ground state with traps
contain non-Abelian quasiparticles, the resulting degeneracy is robust against
any local perturbations.  So the system is a stable gapped quantum system.  However,
for such a system, $H_{N_k}$ and $H_{N_{k+1}}$ are not connected via gLU transformations,
hence it is not a gapped quantum liquid system.

Now we can define topological order (or different phases of 
topologically ordered states):
\begin{defn}\label{def:topo} \textbf{Topological order}\\
\noindent The topological orders are stable gapped quantum liquid phases.
\label{def:topo} 
\end{defn}
\noindent

We remark that we in fact define different topological orders as different equivalent classes.
One of these equivalent classes represents the trivial (topological) order.
In Definition~\ref{def:topo}, we put trivial and non-topological order together to have a simple definition. 
This is similar to symmetry transformations, which usually include both trivial and non-trivial transformations, so that we can say symmetry transformations form a group. Similarly, if we include the trivial one, then we can say that topological orders form a monoid under the stacking operation~\cite{KW1458}.

There are also unstable gapped quantum liquid systems. They can be defined via
the definition of
\begin{defn} \textbf{First order phase transition for gapped quantum liquid systems}\\
A deformation of a gapped quantum liquid system experiences a first order phase
transition if the Hamiltonian remains gapped along the deformation path and if
the ground-state degeneracy at a point on the deformation path is different
from its neighbours.  That point is the transition point of the first order
phase transition.  \end{defn} 
\noindent The first order phase-transition point
is also an unstable gapped quantum liquid system.  Physically and generically,
an unstable gapped quantum liquid system is a system with accidental degenerate
ground states.  

From the above discussions, we see that topological orders are the universality
classes of stable gapped quantum liquid systems that are separated by gapless
quantum systems or unstable gapped quantum systems.  Moving from one
universality class to another universality class by passing through a gapless
system corresponds to a continuous phase transition.  Moving from one
universality class to another universality class by passing through an unstable
gapped system corresponds to a first order phase transition.

\subsection{Gapped quantum liquid}

We would like to emphasize that the topological order is a notion of
universality classes of local Hamiltonians (or more precisely, gapped quantum
systems). In the following, we will introduce the universality classes of
many-body wavefunctions.  We can also use the universality classes of
many-body wavefunctions to understand topological orders.

\begin{defn}
\textbf{Gapped quantum state}\\
A gapped quantum system is defined by a sequence of Hamiltonians $\{H_{N_k}\}$.
Let $\mathcal{V}_{N_k}$ be the  ground-state subspace of $H_{N_k}$.
The sequence of ground-state subspaces $\{\mathcal{V}_{N_k}\}$
is referred as a gapped quantum state. 
\end{defn} \noindent
Note that a gapped quantum state is not described by a single wavefunction,
but by a sequence of ground-state subspaces $\{\mathcal{V}_{N_k}\}$.
Similarly,
\begin{defn}
\textbf{Gapped quantum liquid}\\
The sequence of ground-state subspaces $\{\mathcal{V}_{N_k}\}$ of a gapped quantum
liquid system defined by $\{H_{N_k}\}$ is referred to as a gapped quantum liquid. 
\end{defn} \noindent

Now we are ready to define
\begin{defn}
\textbf{Gapped quantum liquid phase and topologically ordered phase}\\ Two gapped
quantum liquids, defined by two sequences of ground-state subspaces
$\{\mathcal{V}_{N_k}\}$ and $\{\mathcal{V}'_{N'_k}\}$ (on space with no boundary), are
equivalent if they can be connected via gLU transformations, i.e.\ we can map
$\mathcal{V}_{N_k}$ into $\mathcal{V}'_{N'_k}$ and map $\mathcal{V}'_{N'_k}$ into $\mathcal{V}_{N_k}$ via
gLU transformations (assuming $N_k\sim N'_k$).  The equivalence classes of
gapped quantum liquids are gapped quantum liquid phases.  The equivalence classes of
stable gapped quantum liquids are topologically ordered phases.
\end{defn} \noindent
In the next section, we will show that gapped liquid phases contain both
symmetry-breaking phases and topologically ordered phases.

\begin{figure}[tb] 
\begin{center} 
\includegraphics[scale=0.2]{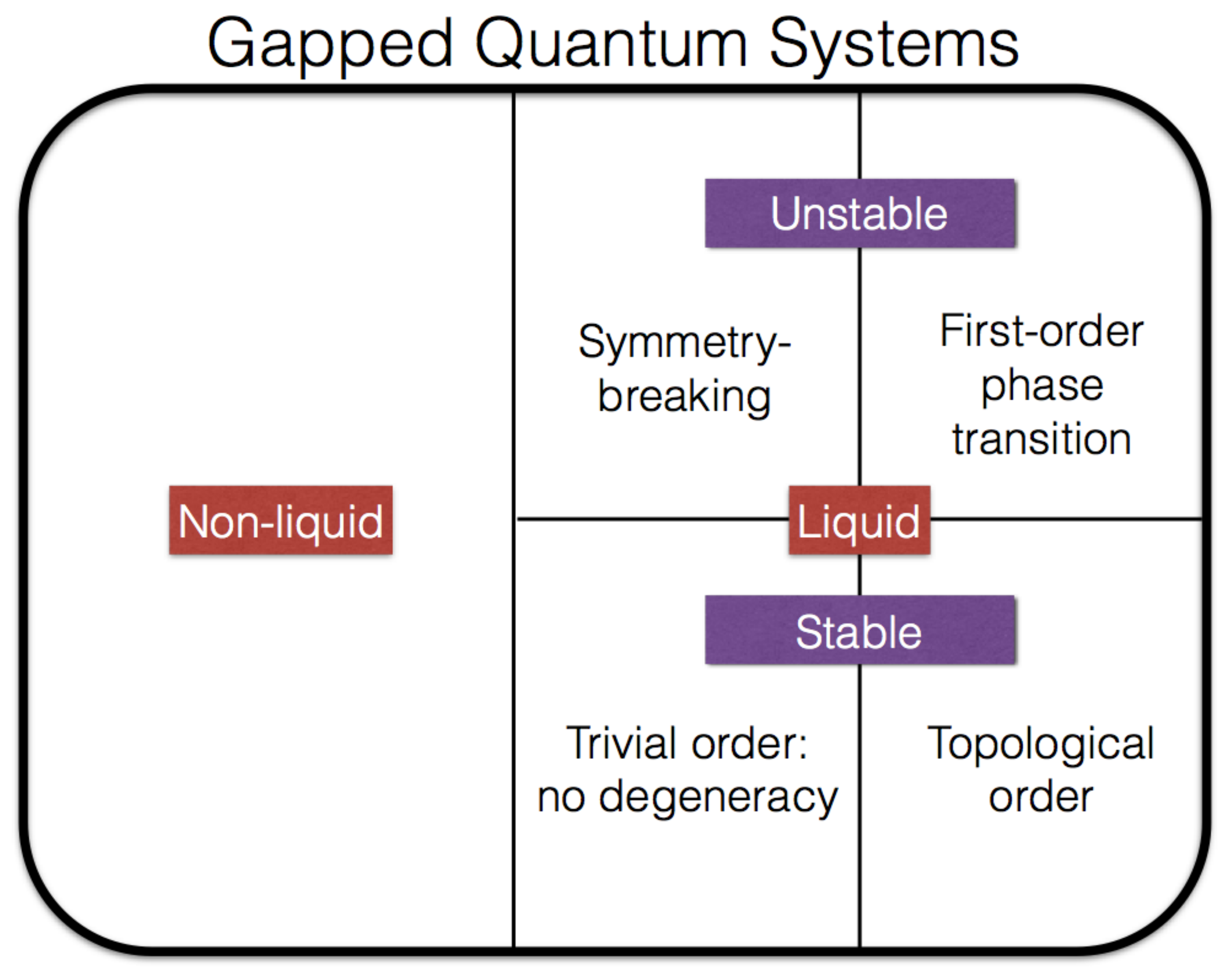} \end{center}
\caption{
Summary of gapped quantum systems: gapped quantum systems
include gapped quantum liquid systems, and systems that are not liquid (nonliquid).
For gapped quantum liquids, there are stable systems (including the trivial systems
given by e.g. the Hamiltonian $H^\text{non-liquid}_{N_k}$ and the topologically ordered
systems) and unstable systems (including symmetry-breaking systems and first-order
phase transitions).
} 
\label{Liquid} 
\end{figure}

We summarize the different kinds of gapped quantum systems in Fig.~\ref{Liquid}.

\section{Local unitary transformations and 
symmetry-breaking order}

To study the universality classes of many-body wave functions, a natural idea
is from the LU transformations~\cite{CGW1038}.  In this section we will analyze
the classes of wave functions under LU transformations, or more generally, gLU
transformation.

As discussed above, the gLU transformations define an equivalence relation among
many-body ground-state subspaces.  The equivalence classes defined by such an
equivalence relation will be called the gLU classes.  The gLU classes of gapped
quantum liquids correspond to gapped quantum liquid phases.

We now ask the following question.
\begin{ques}
Since the definition of the gLU classes does not require symmetry, then
do the gLU classes of gapped quantum liquid have a one-to-one correspondence
with topological orders (as defined in Defition~\ref{def:topo})?
\end{ques}
We will show that the answer is no, i.e.\ there are \emph{unstable} gapped quantum
liquids. Only the gLU classes for \emph{stable} gapped quantum liquids have a
one-to-one correspondence with topological orders.

\subsection{Symmetry-breaking orders}

An example of  unstable gapped quantum liquids is given by symmetry breaking
states.  Those unstable gapped quantum liquids are in a different gLU class
from the trivial phase, and thus are non-trivial gapped quantum liquid phases.

Let us consider an example of the unstable gapped quantum liquids, the 1D
transverse Ising model with the Hamiltonian (with periodic boundary condition)
\begin{equation}
H^\text{tIsing}_{N_k}(B)=-\sum_{i=1}^{N_k}Z_iZ_{i+1}+B\sum_{i=1}^{N_k}X_i,
\end{equation}
where $Z_i$ and $X_i$ are the Pauli $Z/X$ operators
acting on the $i$th qubit.
The Hamiltonian $H^\text{tIsing}_{N_k}(B)$ has a $\mathbb{Z}_2$ symmetry, which is given by 
$\prod_{i=1}^{N_k}X_i$.
The gapped ground states are non-degenerate for $B>1$. For $0\leq B<1$, the
gapped ground states are two-fold degenerate.  The degeneracy is unstable
against perturbation that breaks the $\mathbb{Z}_2$ symmetry.

The phase for $B>1$ is a trivial gapped liquid phase.  The phase for $0<B<1$ is
a non-trivial gapped liquid phase.  This due to a very simple reason: the two
phases have different group state degeneracy, and the ground-state degeneracy
is an gLU invariant.  Gapped quantum liquids with different ground-state degeneracy always belong to different gapped liquid phases.

Now, let us make a more non-trivial comparison. Here we view
$H^\text{tIsing}_{N_k}(B)$ (with $0<B<1$) as a gapped quantum system (rather
than a gapped quantum liquid system).  We compare it with another gapped
quantum system $H^\text{non-liquid}_{N_k}$ (see \eqn{Hnonliquid}) discussed
before.  Both gapped systems have two-fold degenerate ground states.  Do the
two systems belong to the same gapped quantum phase (as defined in Definition
\ref{gPhase})?

Consider $H^\text{tIsing}_{N_k}(B)$ for any $0<B<1$ and any size $N_k<\infty$. The (symmetric) exact ground
state $\ket{\Psi_+(B)}$ is an adiabatic continuation of the GHZ state
\begin{equation}
\ket{GHZ_+}=\frac{1}{\sqrt{2}}(\ket{0}^{\otimes N_k}+\ket{1}^{\otimes N_k}),
\end{equation}
i.e.\  $\ket{\Psi_+(B)}$ is in the same gLU class of $\ket{GHZ_+}$. 
There is another state $\ket{\Psi_-(B)}$ below the energy window $\Delta$ which is an adiabatic continuation of the state
\begin{equation}
\ket{GHZ_-}=\frac{1}{\sqrt{2}}(\ket{0}^{\otimes N_k}-\ket{1}^{\otimes N_k}).
\end{equation}
The energy splitting of $\ket{\Psi_+(B)}$ and $\ket{\Psi_-(B)}$ approaches zero as $N_k \to \infty$.

However, we know that the GHZ state $\ket{GHZ_+}$ (hence $\ket{\Psi_+(B)}$) and
the product state $\ket{0}^{\otimes N_k}$ belong to two different gLU classes.
Both states are regarded to have the same trivial topological order.  So gLU
transformations assign GHZ states, or symmetry-breaking many-body wave
functions, to non-trivial classes.  Therefore by studying the gLU classes of
gapped quantum liquids, we can study both the topologically ordered states and
the symmetry-breaking states.

To be more precise, the ground-state subspace of $H^\text{tIsing}_{N_k}(B)$
($0<B<1$) contain non-trivial GHZ states.  On the other hand, the ground-state
subspace of $H^\text{non-liquid}_{N_k}$ contain only product states. There is
no GHZ states.  That make the two systems  $H^\text{tIsing}_{N_k}(B)$ and
$H^\text{non-liquid}_{N_k}$ to belong to two different gapped quantum phases,
even though the two systems have the same ground-state degeneracy.

We can now define:
\begin{defn} \textbf{Gapped symmetry-breaking quantum system}\\
A gapped symmetry-breaking system is a gapped quantum liquid system 
with certain symmetry and degenerate ground states, where
the symmetric ground states have the GHZ form of entanglement.
\end{defn} \noindent

We remark that the ground-state subspace of a gapped symmetry-breaking quantum
system is a ``classical" error-correcting code with macroscopic distance,
correcting errors that does not break the symmetry. This is to say, for any
orthonormal basis $\{\ket{\Phi_i}\}$ of the ground-state subspace, for any local
operator $M_s$ that does not break symmetry, we have 
\begin{equation}
\bra{\Phi_i}M_s\ket{\Phi_j}=C_{M_s}\delta_{ij},
\end{equation}
where $M_s$ is a constant that only depends on $M_s$.

Here by ``classical" we mean the following. 
For the ground-state subspace,
there exists a basis $\{\ket{\Phi_i}\}$ 
that is connected by symmetry. 
In this basis, the ground-state subspace is a classical error-correcting code of macroscopic distance,
in the sense that for any local operator $M$, we have 
\begin{equation}
\label{eq:ccode}
\bra{\Phi_i}M\ket{\Phi_j}=0,\ i\neq j.
\end{equation}
Notice that Eq~\eqref{eq:ccode} does not contain the coherence condition for
$i=j$, which is the requirement to make the subspace a `quantum' code.

The transverse Ising mode is an example of such a special case with
$\mathbb{Z}_2$ symmetry. The basis that is connected by the $\mathbb{Z}_2$
symmetry are $\ket{\Psi_\pm(B)}$.  And it is obvious that
$\bra{\Psi_+(B)}M\ket{\Psi_-(B)}=0,\ i\neq j$.

\begin{figure}[tb] 
\begin{center} 
\includegraphics[scale=0.2]{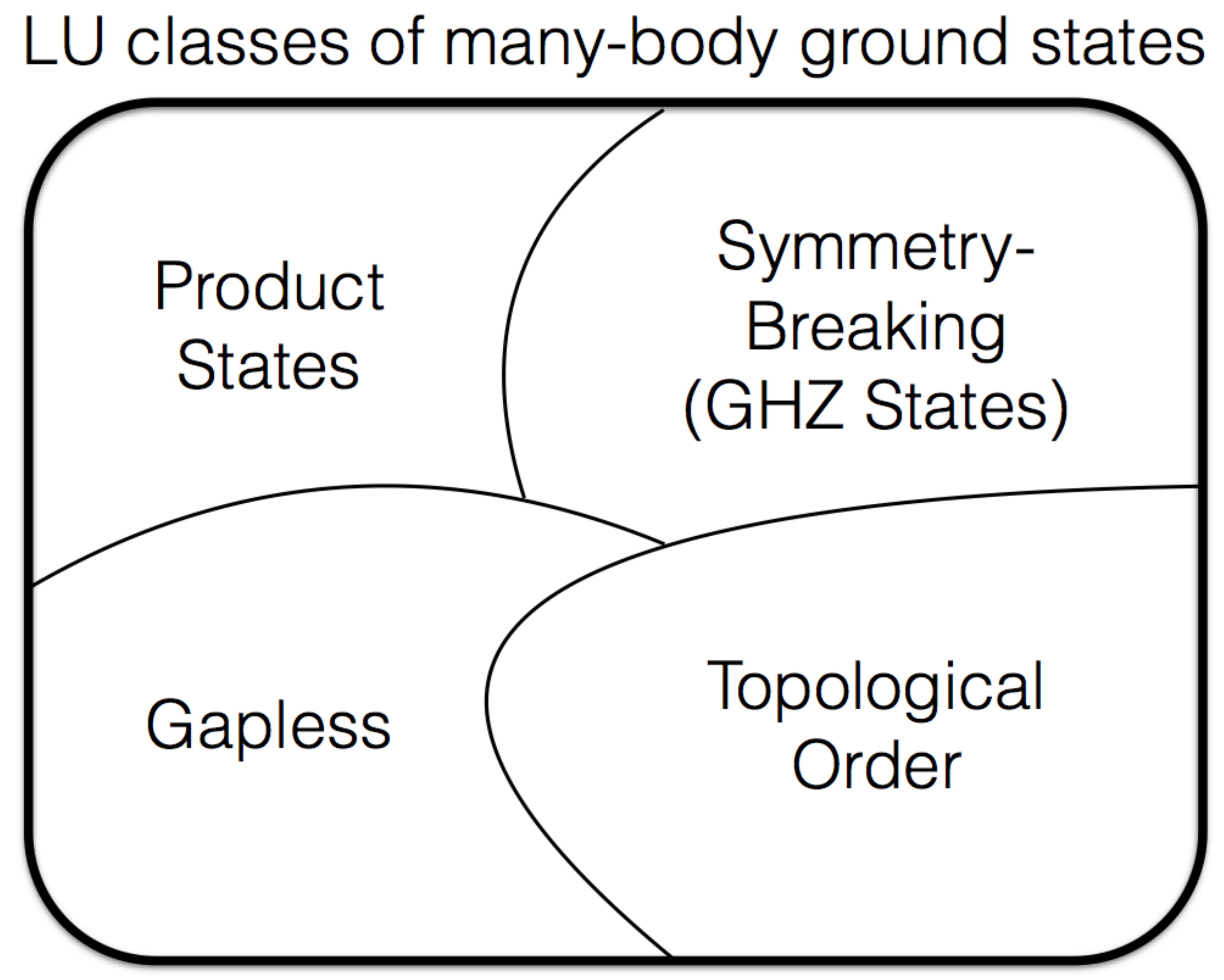} \end{center}
\caption{
LU classes for ground states (many-body wave functions) of local Hamiltonians.
} 
\label{LU1} 
\end{figure}

We have now shown that gapped liquid phases also contain
symmetry-breaking phases. We summarize 
the LU classes for ground states of local Hamiltonians
in Fig.~\ref{LU1}.

\subsection{Topological entanglement entropy}

Topological entanglement entropy is an invariant of gLU
transformations~\cite{LW0605}.  This allows us to use topological entanglement
entropy to detect if a gapped quantum liquid belongs to a non-trivial gLU class
or not, hence we can study both topological orders and symmetry-breaking
orders.

\begin{figure}[tb] 
\begin{center} 
\includegraphics[scale=0.5]{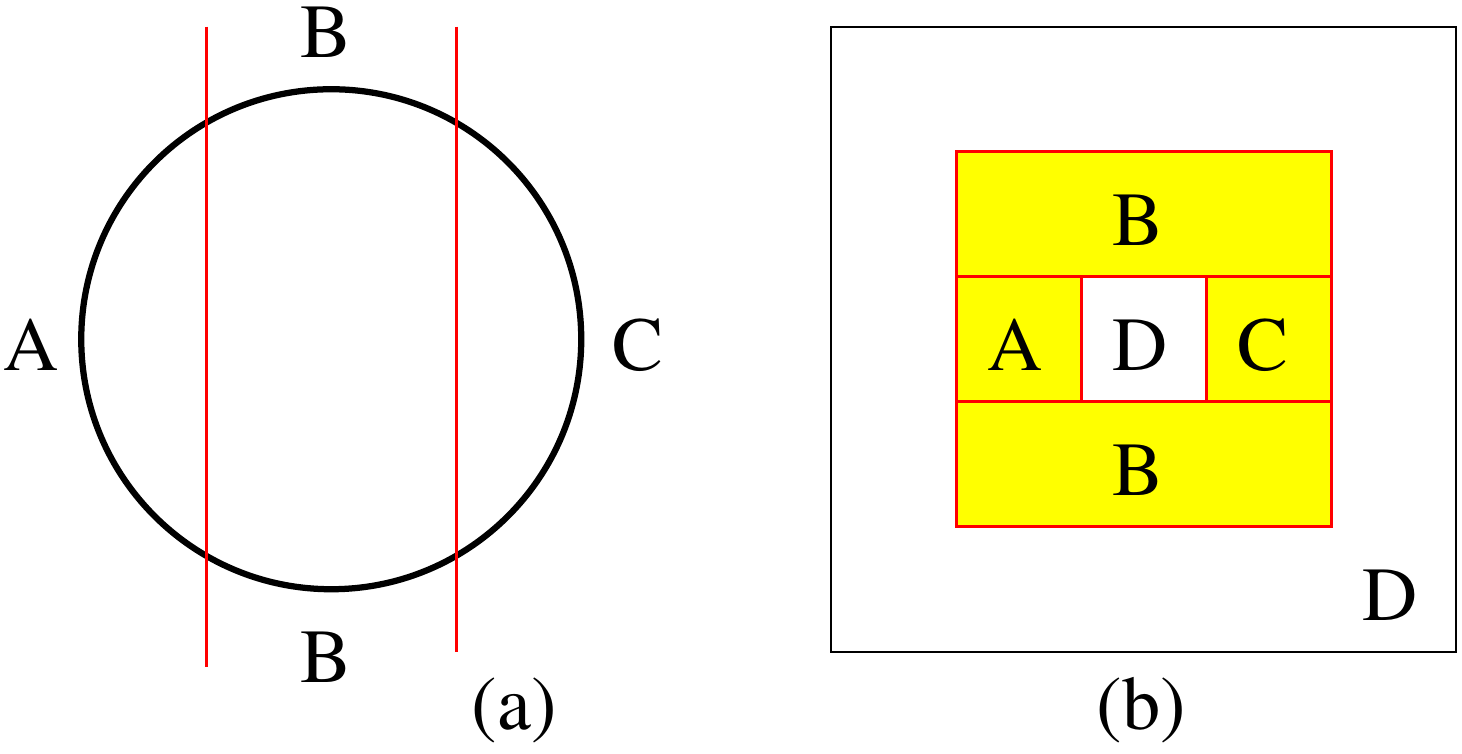} \end{center}
\caption{
(a) Cutting a ring into three parts $A,B,C$.
(b) Cutting a 2D system into four parts $A,B,C,D$.
} 
\label{Cut} 
\end{figure}

Here we define a new type of topological entanglement entropy on a graph with
no boundary (for simplicity we consider the topology of a $D$-dimensional sphere
$\mathcal{S}^D$), by dividing the entire system into \emph{three} non-overlaping
parts $A,B,C$.
\begin{defn}
The tri-topological entanglement entropy on $\mathcal{S}^D$ is given by 
\begin{equation}
\label{eq:tritopo}
S^\text{tri}_\text{topo}=S(AB)+S(BC)-S(B)-S(ABC),
\end{equation}
where the parts $A,B,C$ are illustrated in Fig.~\ref{Cut}a for the case of 
$d=1$. And $S(*)$ is the von Neumann entropy of the reduced density 
matrix of the part $*$.
\end{defn} \noindent
We note that, in the original definition of topological entropy (denoted as
$S^\text{qua}_\text{topo}$), the entire system is cut into \emph{four} unconnected
non-loverlaping pieces $A,B,C,D$~\cite{LW0605}.  $S^\text{qua}_\text{topo}$ is defined
on the part of the system $A,B,C$ with nontrivial topology (see Fig.~\ref{Cut}b):
\begin{equation}
S^\text{qua}_\text{topo}=S(AB)+S(BC)-S(B)-S(ABC).
\end{equation}

We remark that the topological entanglement entropy given by $S^\text{qua}_\text{topo}$ 
has the reversed sign compared to the original definition~\cite{LW0605,KP0604}.
Also, both $S^\text{tri}_\text{topo}$ and $S^\text{qua}_\text{topo}$ are 
guaranteed to be non-negative due to the strong subadditivity of quantum entropy~\cite{lieb2002proof}.

Notice that for product states $S^\text{tri}_\text{topo}=S^\text{qua}_\text{topo}=0$.  However,
we have 
$S^\text{tri}_\text{topo}(\ket{GHZ_+})=1$.  And one expects the same $S^\text{tri}_\text{topo}=1$
for $\ket{\Psi_+(B)}$ when $N_k \to \infty$ and $0<B<1$, which is not an ideal
GHZ state.  That is, $S^\text{tri}_\text{topo}$ can detect the states with GHZ form of entanglement. Since the product state $\ket{0}^{\otimes N_k}$ has zero
tri-topological entropy, it belongs to a different gLU class. 

\begin{figure}[tbh!] 
\begin{center} 
\includegraphics[scale=0.3]{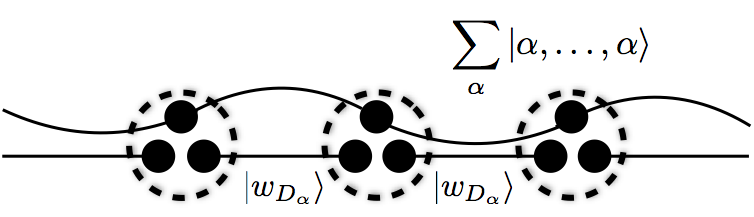} \end{center}
\caption{Isometric form for MPS with GHZ form of entanglement~\cite{SPC5139}
.} 
\label{GHZ} 
\end{figure}

The long-range correlation of 
the wavefunction may be better
seen by the matrix product state (MPS) renormalization group, where the
isometric form is given by~\cite{SPC5139}
\begin{equation}
\sum_{\alpha}\ket{\alpha,\ldots,\alpha}\otimes\ket{\omega_{D_{\alpha}}}^{\otimes N_k},
\end{equation}
as illustrated in Fig.~\ref{GHZ}. Here the bond state is
\begin{equation}
\ket{\omega_{D_{\alpha}}}=\frac{1}{\sqrt{D_{\alpha}}}\sum_{j=1}^{D_{\alpha}}\ket{jj},
\end{equation}
and the bond dimension $D_{\alpha}$ can couple to the value of $\alpha$.

The term $\sum_{\alpha}\ket{\alpha,\ldots,\alpha}$ captures the GHZ form of entanglement which contributes to $S^\text{tri}_\text{topo}$, and the term $\ket{\omega_{D_{\alpha}}}^{\otimes N_k}$ is short-range entangled part which has $S^\text{tri}_\text{topo}=0$.

Notice that, if the state of the system $ABC$ is a pure state, we have 
\begin{eqnarray}
S(AB)&=&S(C),\nonumber\\
S(BC)&=&S(A),\nonumber\\
 S(B)&=&S(AC),\nonumber\\
 S(ABC)&=&0.
\end{eqnarray}
In this case the right hand side of Eq.~\eqref{eq:tritopo} becomes the mutual
information between the parts $A$ and $C$, i.e.\
\begin{equation}
I(A:C)=S(A)+S(C)-S(AC).
\end{equation}
This gives an alternative explanation that a nonzero mutual information of two 
disconnected parts of a pure state indicates long-range correlation, as 
is observed in~\Ref~\cite{PhysRevLett.102.170602}.

However, in general, a nonzero $I(A:C)$ does not indicate a GHZ form of entanglement.
For instance, the state
\begin{equation}
\rho=\frac{1}{2}(\ket{\bar{0}}\bra{\bar{0}}+\ket{\bar{1}}\bra{\bar{1}}),
\end{equation}
where $\ket{\bar{0}}=\ket{0}^{\otimes N_k}$ and $\ket{\bar{1}}=\ket{1}^{\otimes N_k}$,
contains only classical correlation but no entanglement.
For $\rho$, we have $I(A:C)=1$ but $S^{tri}_{topo}=0$.

We remark that, $S^\text{tri}_\text{topo}$ is evaluated for a quantum 
state of the region $ABC$.
For finite systems, we will focus on the value of $S^\text{tri}_\text{topo}$
on the exact ground state, which is non-degenerate and does not break 
any symmetry. We refer to~\cite{furukawa2006systematic,furukawa2007reduced,
cheong2009correlation,henley2014density} for
some other approaches proposed to detect orders of 
the systems based on density matrices.

As an example to demonstrate the use of $S^\text{tri}_\text{topo}$ to determine the
quantum phase transitions, we calculate $S^\text{tri}_\text{topo}$ for the ground state
of the transverse Ising model. We rescale the Ising Hamiltonian $H(B)$ as
\begin{equation}
\label{eq:Halpha}
H(\alpha)=-\cos\alpha\sum_{i}Z_iZ_{i+1}+\sin\alpha\sum_i X_i,
\end{equation} 
where $\alpha\in[0,\pi/2]$. 

\begin{figure}[tbh!] 
\begin{center} 
\includegraphics[scale=0.3]{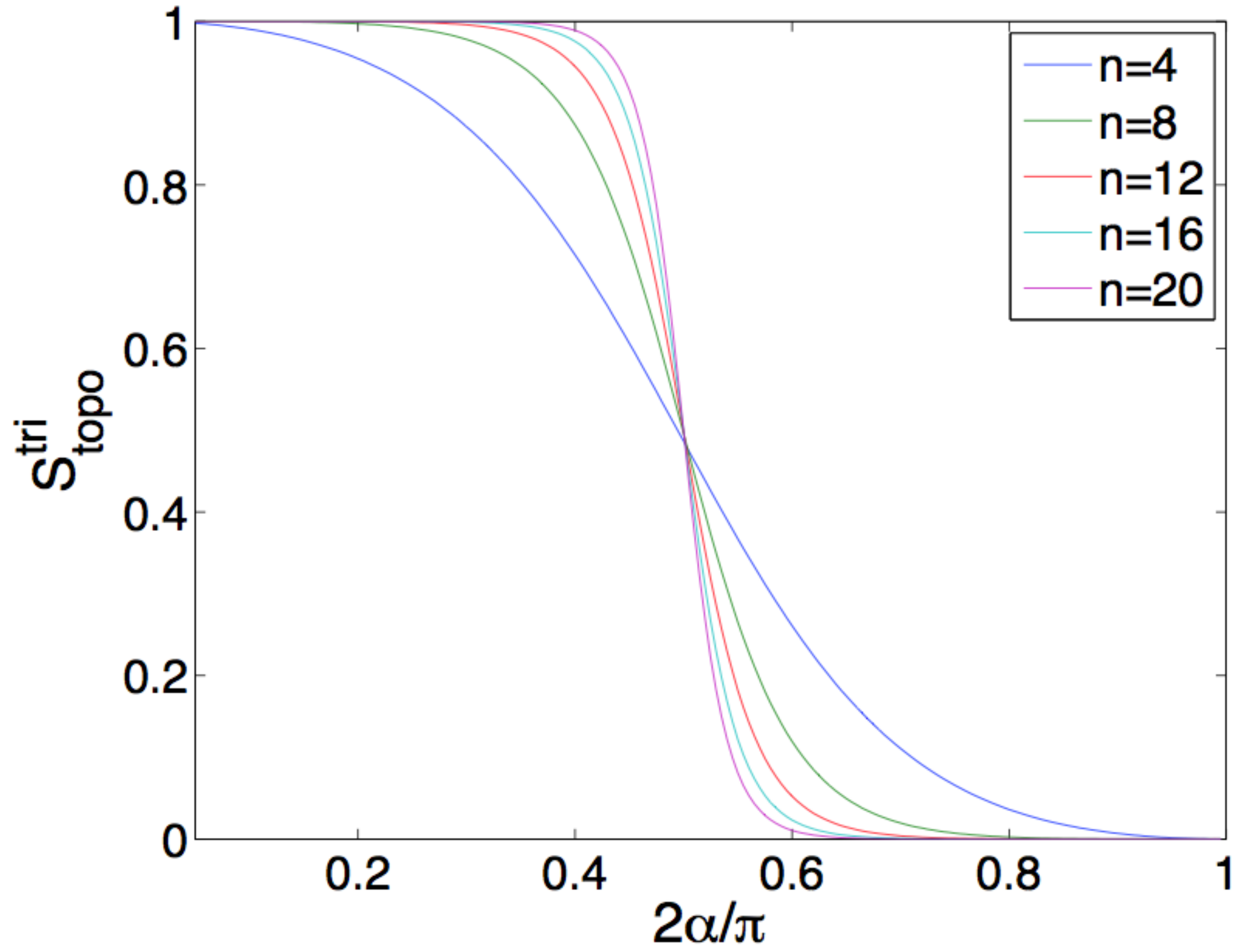} 
\end{center}
\caption{$S^\text{tri}_\text{topo}$ for the transverse Ising model. The horizontal 
axis is the angle $2\alpha/\pi$ for the Hamiltonian $H(\alpha)$ as given in Eq.~\eqref{eq:Halpha}.
A similar result is presented in ~\Ref~\cite{MaxEnt}, from a different viewpoint.} 
\label{Ising} 
\end{figure}

We choose the area $A,C$ and each connected component of the area $B$ to have $1,2,3,4,5$ qubits, so we compute $S^\text{tri}_\text{topo}$ for total $n=4,8,12,16,20$ qubits. The results are shown in Fig.~\ref{Ising}.
The five curves intersect at $2\alpha/\pi=1/2$, i.e.\ $\alpha=\pi/4$, which corresponds to the
well-known phase transition at $B=1$.

We emphasize that for both the symmetric ground states of the symmetry breaking
phase and the trivial phase, the tri-topological entanglement entropy
$S^\text{tri}_\text{topo}$ is quantized.  In this sense, it is similar to the topological
entanglement entropy $S^\text{qua}_\text{topo}$ of topologically ordered ground states.
However, the symmetry-breaking classes are quite different from topological
classes: $S^\text{tri}_\text{topo}\neq 0$ and $S^\text{qua}_\text{topo} = 0$ for symmetry-breaking
classes, while $S^\text{tri}_\text{topo}= 0$ and $S^\text{qua}_\text{topo} \neq 0$ for topological
classes (with non-trivial topological excitations\cite{KW1458}).
We see that, for symmetry-breaking classes, the original definition
$S^\text{qua}_\text{topo}$ fails to detect different gLU classes. This is because that
for symmetry-breaking class, the information of the non-trivial entanglement
is only contained in the wave function for the entire system.  Reduced density
matrices of any part of the system do not contain that information.

\begin{figure}[tbh!] 
\begin{center} 
\includegraphics[scale=0.3]{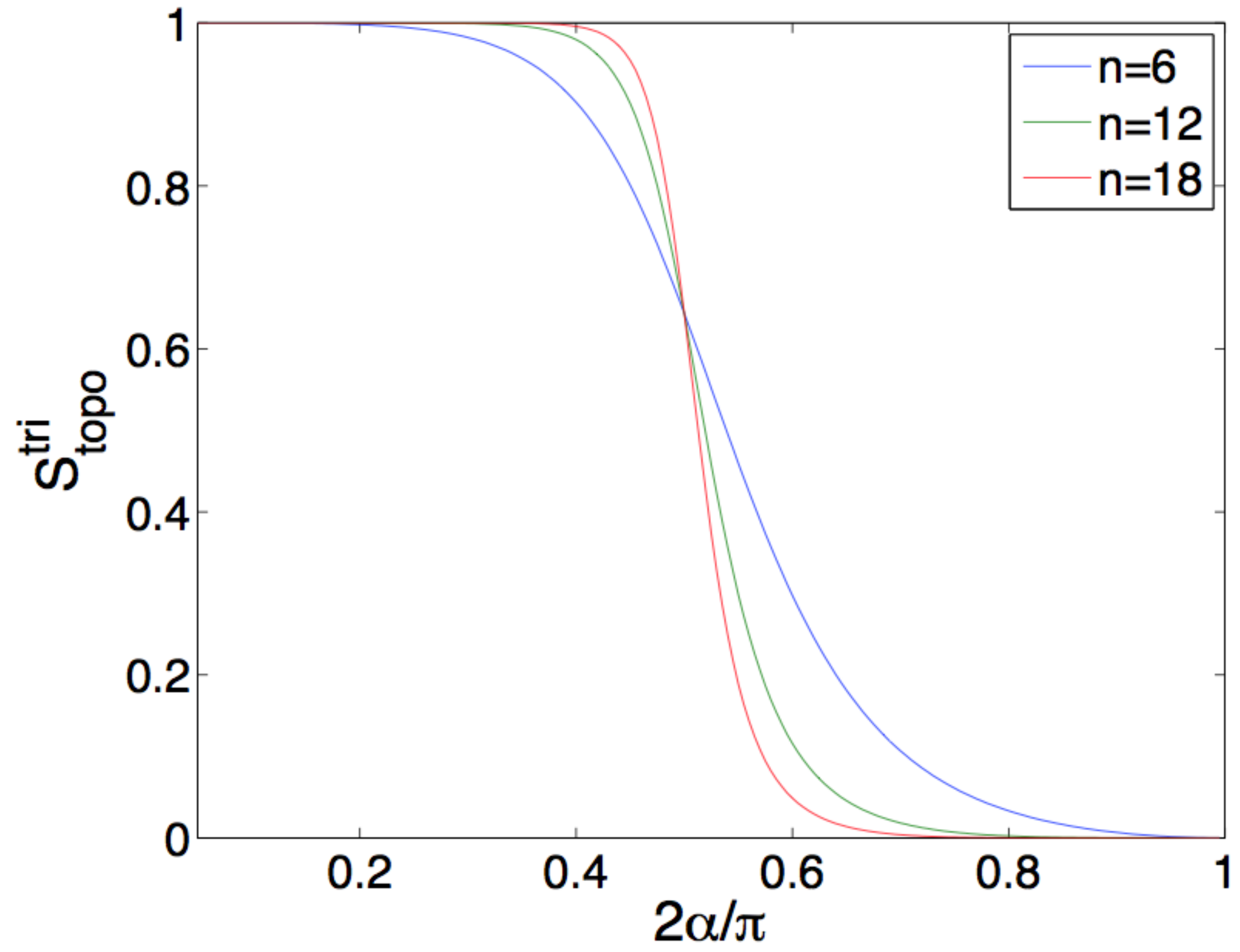} 
\end{center}
\caption{$S^\text{tri}_\text{topo}$ for the transverse Ising model. The ratio $r=2:1$, and the system sizes are $n=6,12,18$ qubits.} 
\label{Ising1} 
\end{figure}

We remark that, at the transition point $\alpha=\pi/4$ where the system is gapless, the five curves
in Fig.~\ref{Ising} intersect at $S^\text{tri}_\text{topo}\sim 0.5$. However, this value of  $S^\text{tri}_\text{topo}$ is not a constant, which depends on the shape of the areas $A,B,C$. For instance,
if we choose the ratio 
\begin{equation}
r=\frac{\#\ \text{in each of the area A,C}}
{\#\ \text{in each connected component of $B$}}
\end{equation}
$2:1$, where $\#$ means the number of qubits, and to have $1,2,3$ qubits for each connected component of the area $B$, then we can compute $S^\text{tri}_\text{topo}$ for total $n=6,12,18$ qubits, as shown in Fig.~\ref{Ising1}. 
This ratio dependence is typical in critical systems~\cite{PhysRevLett.102.170602,calabrese2009entanglement}, and our results are consistent with the conformal field theory predictions~\cite{PhysRevLett.102.170602,de2015entanglement}.

\section{Stochastic local transformations and long-range entanglement}

We have seen that the non-trivial equivalence classes of many-body wave
functions under the gLU transformations contain both topologically ordered
phases and symmetry-breaking phases (described by the symmetric many-body wave
functions with GHZ form of entanglement).  In this section, we will introduce
the generalized stochastic local (gSL) transformations, which are local
invertible transformations that are not necessarily unitary.  The term
`stochastic' means that these transformations can be realized by generalized
local measurements with finite probability of success~\cite{bennett2000exact}. 

We show that the many-body wave functions for symmetry-breaking phases (i.e.\ the
states of GHZ form of entanglement) are convertible to the product states under
the gSL transformations with a finite probability, but in contrast the
topological ordered states are not. This allows us to give a new definition of
long-range entanglement under which only topologically ordered states are
long-range entangled.  We further show that the topological orders are stable
against small stochastic local transformations, while the symmetry-breaking
orders are not.

\subsection{Stochastic local transformations}

The idea for using gSL transformations is simple.  The topologically
\emph{stable} degenerate ground states for a topologically ordered system is
not only stable under real-time evolutions (which are described by gLU
transformations), they are also stable and are the fixed points under
imaginary-time evolutions.  The imaginary-time evolutions of the ground states
are given by the gSL transformations (or local non-unitary transformations),
therefore the topological orders are robust under (small) gSL transformations.

On the other hand, the states of GHZ form of entanglement are not robust under
small gSL transformations, and can be converted into product states with a
finite probability.  Thus, there is no emergence of unitarity for symmetry-
breaking states.

To define gSL transformations, we start from reviewing the most general form of
quantum operations (also known as quantum channels), which are
complete-positive trace-preserving maps~\cite{nielsen2010quantum}. A quantum
operation $\mathcal{E}$ acting on any density matrix $\rho$ has the form
\begin{equation}
\mathcal{E}(\rho)=\sum_{k=1}^r A_k\rho A_k^{\dag},
\end{equation} 
with 
\begin{equation}
\sum_{k=1}^r A_k^{\dag}A_k=I,
\end{equation} 
where $I$ is the identity operator.

The operators $A_k$ are called Kraus operators
of $\rho$ and satisfies
\begin{equation}
A_k^{\dag}A_k\leq I.
\end{equation}
This means that the operation $A_k\rho A_k^{\dag}$
can be realized with probability $\Tr(A_k\rho A_k^{\dag})$
for a normalized state $\Tr\rho=1$. In the following 
we will drop the label $k$ for the measurement outcome. 

We will now definite gSL transformations along a similar line 
as the definition of gLU transformations. Let us first define
a layer of SL transformation that has a form
\begin{equation*}
 W_{pwl}= \prod_{i} W^i
\end{equation*}
where $\{ W^i \}$ is a set of invertible operators that act on non-overlapping
regions, and each $W^i$ satisfies
\begin{equation}
\label{eq:trace}
W^{i\dag}W^i\leq I.
\end{equation}
The size of each region is less than a finite number $l$. The
invertible operator $W_{pwl}$ defined in this way is called a layer of
piecewise stochastic local transformation with a range $l$.  

A stochastic local (SL) transformation is then given by a finite layers of piecewise
local invertible transformation:
\begin{equation*}
 W^M_{circ}= W_{pwl}^{(1)} W_{pwl}^{(2)} \cdots W_{pwl}^{(M)}
\end{equation*}
We note that such a transformation does not change the degree of freedom of
the state.

Similarly to the gLU transformations, we can also have a transformation that can change the degree of freedom of
the state, by a tensor product of the state with another product state $
 \ket{\Psi} \to 
\Big(\otimes_i\ket{\psi_i}\Big) 
\otimes \ket{\Psi}
$,
where $\ket{\psi_i}$ is the wave function for the $i^\text{th}$ qubit.  A finite
combination of the above two types of transformations is 
then a generalized stochastic local (gSL) transformation.

We remark that, despite the simple idea similar to the gLU transformations, gSL transformations are more subtle to deal with. 
First of all, notice that gSL transformations do not preserve the norm of quantum states (i.e.\ not trace-preserving, as given by Eq.~\eqref{eq:trace}).
Furthermore, as we are dealing with thermodynamic limit ($N_k\rightarrow\infty$), we are applying gSL transformations on a system of infinite dimensional Hilbert space. In this case, even if each $W^i$ is invertible, $W_{pwl}= \prod_{i} W^i$ may be non-invertible due to 
the thermodynamic limit. We will discuss these issues in more detail
in the next subsection.

\subsection{Short-range entanglement and symmetry-breaking orders}
It is known in fact that the SL convertibility in 
infinite dimensional systems is subtle, and to avoid
technical difficulties dealing with the infinite dimensional Hilbert space,
we would instead start from borrowing the idea in \Ref~\cite{owari2008varepsilon} to use $\epsilon$-convertibility instead to talk about the exact convertibility of states under gSL. For simplicity
we will omit the notation ``$\epsilon$" and still name it ``gSL convertibility".

\begin{defn}\textbf{Convertibility by gSL transformation}\\
\label{def:gSLc}
We say that $\ket{\Psi}$ is convertible to $\ket{\Phi}$ by a gSL transformation, if for any $\epsilon>0$, there exists an integer $N$, a probability $0<p<1$, and gSL transformations $W_{N_k}$, such that for any $N_k>N$, $W_{N_k}$ satisfy the condition 
\begin{equation}
\left\|\frac{W_{N_k}(\ket{\Psi}\bra{\Psi})W^{\dag}_{N_k}}{\Tr\left(W_{N_k}(\ket{\Psi}\bra{\Psi})W^{\dag}_{N_k}\right)}
-\frac{\ket{\Phi}\bra{\Phi}}{\Tr(\ket{\Phi}\bra{\Phi})}\right\|_{\text{tr}}<\epsilon,
\end{equation}
where $\|\cdot\|_{\text{tr}}$ is the trace norm and
\begin{equation}
\label{eq:prob}
\frac{\Tr(W_{N_k}\ket{\Psi}\bra{\Psi}W^{\dag}_{N_k})}{\Tr(\ket{\Psi}\bra{\Psi})}>p.
\end{equation}
\end{defn}

The idea underlying Definition~\ref{def:gSLc} is that $\ket{\Psi}$ can be transformed to any neighbourhood of $\ket{\Phi}$, though not $\ket{\Phi}$ itself, and these neighbourhood states become indistinguishable from $\ket{\Phi}$ in the thermodynamic limit.

Using the idea of gSL transformations, we can have a new definition for short-range
and long-range entanglement (`new' in a sense that the previous definition 
was given by gLU transformations).
\begin{defn} \textbf{Short/Long-Range Entanglement}\\
A state is short-range entangled (SRE) if it is convertible to a product
state by a gSL transformation. Otherwise the state is long-range entangled (LRE).
\end{defn} \noindent

Under this new definition, the states which can be transformed to
product states by gLU transformations are SRE. However, the
SRE states under gSL transformations will also include some
of the states that cannot be transformed to 
product states by gLU transformations. 

As an example, the state 
\begin{equation}
\ket{GHZ_+(a)}=a\ket{0}^{\otimes N_k}+b\ket{1}^{\otimes N_k}
\end{equation} 
with $|a|^2+|b|^2=1$
cannot be transformed 
to product states under gLU transformations.
However if one allows gSL transformations,
then all the $\ket{GHZ_+(a)}$ are convertible to $\ket{GHZ_+(1)}$, i.e.\ the
product state $\ket{0}^{\otimes N_k}$. To see
this, one only needs to apply the gSL transformation
\begin{equation}
\label{eq:SLeg}
W_{N_k}=\prod_{i=1}^{N_k} O_i,
\end{equation}
where $O_i$ is the invertible operator
\begin{equation}
\begin{pmatrix}
1 & 0 \\ 0 & \gamma
\end{pmatrix}
\end{equation}
acting on the $i$ the qubit, and $0<\gamma<1$.
And we have 
\begin{equation}
\begin{pmatrix}
1 & 0 \\ 0 & \gamma
\end{pmatrix}^{\dag}\begin{pmatrix}
1 & 0 \\ 0 & \gamma
\end{pmatrix}\leq \begin{pmatrix}
1 & 0 \\ 0 & 1
\end{pmatrix}=I.
\end{equation}

That is
\begin{eqnarray}
\label{eq:GHZc}
W_{N_k}\ket{GHZ_+(a)}&=&
a\ket{0}^{\otimes N_k}+b\gamma^{N_k}\ket{1}^{\otimes N_k}\nonumber\\
&=&\ket{\mathcal{V}_{N_k}}.
\end{eqnarray}
Obviously, the right hand side of Eq.~\eqref{eq:GHZc}
can be arbitrarily close to the product state $\ket{0}^{\otimes N_k}$ 
as long as $N_k$ is large enough. Furthermore, $\Tr(\ket{\mathcal{V}_{N_k}}\bra{\mathcal{V}_{N_k}})>|a|^2$
for any $N_k$.
Therefore, according to Definition~\ref{def:gSLc}, $\ket{GHZ_+(a)}$ 
is convertible to the product state $\ket{0}^{\otimes N_k}$ by the gSL transformation $W_{N_k}$.

As another example, we can see how to convert a ground state of any 1D gapped quantum liquid to a product state by gSL transformations. Hence there is no long-range entangled states (i.e.\ no topological order) in 1D systems. We again use the MPS isometric form~\cite{SPC5139}
\begin{equation}
\sum_{\alpha}\ket{\alpha,\ldots,\alpha}\otimes\ket{\omega_{D_{\alpha}}}^{\otimes N_k}
\end{equation} 
This state is the convertible to a product state by gSL transformations via two steps: the first step is an gLU transformation to convert the $\ket{\omega_{D_{\alpha}}}^{\otimes N_k}$ part to a product state and end up with a GHZ state. The the next step is to apply the gSL transformation $W_{N_k}$ as given in Eq.~\eqref{eq:GHZc}, which transforms the GHZ state to a product state with a finite probability.

If $\ket{\Psi}$ is convertible to $\ket{\Phi}$ by a gSL transformation, we write
\begin{equation}
\ket{\Psi}\xrightarrow{\text{gSL}}\ket{\Phi}.
\end{equation}
Notice that $\ket{\Psi}\xrightarrow{\text{gSL}}\ket{\Phi}$ does not mean
$\ket{\Phi}\xrightarrow{\text{gSL}}\ket{\Psi}$. 
For example, while we have
\begin{equation}
\ket{\mathcal{V}_{N_k}}\xrightarrow{\text{gSL}}\ket{0}^{\otimes N_k},
\end{equation}
$\ket{0}^{\otimes N_k}$ is not gSL convertible to 
$\ket{\mathcal{V}_{N_k}}$, where $\ket{\mathcal{V}_{N_k}}$
is given in Eq.~\eqref{eq:GHZc}.

That is, 
the gSL convertibility is not an equivalence relation. 
It instead defines a partial order (in terms of set theory)
on all the quantum states. That is,
if $\ket{\Psi}\xrightarrow{\text{gSL}}\ket{\Phi}$
and $\ket{\Phi}\xrightarrow{\text{gSL}}\ket{\Omega}$,
then $\ket{\Psi}\xrightarrow{\text{gSL}}\ket{\Omega}$.
And there exists $\ket{\Psi}$ and $\ket{\Phi}$
that is not comparable under gSL, i.e.\
neither $\ket{\Psi}$ is gSL convertible to
$\ket{\Phi}$, nor $\ket{\Phi}$ is gSL convertible to
$\ket{\Psi}$. Based on this partial order we can 
further define equivalent classes.

\begin{defn}\textbf{gSL equivalent states}\\
We say that two states $\ket{\Psi}$ and $\ket{\Phi}$
are equivalent under gSL transformations if they
are convertible to each other by gSL transformations.
That is, $\ket{\Psi}\xrightarrow{\text{gSL}}\ket{\Phi}$
and $\ket{\Phi}\xrightarrow{\text{gSL}}\ket{\Psi}$.
\end{defn}

Under this definitions, all the states
$\ket{GHZ_+(a)}$ are in the same gSL class
unless $a=0,1$. The product states
with $a=0,1$ are not in the same gSL class,
but any $\ket{GHZ_+(a)}$ is convertible 
to the product states by gSL transformations.
The converse is not true, that a product state 
is not convertible to $\ket{GHZ_+(a)}$ 
with $a=0,1$ by gSL transformations. 

That is to say, the states with GHZ-form of entanglement
are indeed ``more entangled" than product states, but they are 
``close enough" to produce states under gSL transformations.
Furthermore, the topological entanglement entropy
$S^\text{tri}_\text{topo}$ for these types of states are unstable under 
small gSL transformations.
In this sense, we can still treat the GHZ-form of entanglement
as product states, i.e.\ states with no long-range entanglement.

\subsection{Long-range entanglement and topological order}
\label{sec:LRE}

We can now define topologically ordered states based on gSL 
transformations (notice that Definition~\ref{def:topo} defines topological order
through properties of the Hamiltonian).

\begin{defn} \textbf{Topologically ordered states}\\
Topologically ordered states are LRE gapped quantum liquids.  
In other words, a ground state $\ket{\Psi}$ of a gapped Hamiltonian has a
nontrivial topological order if it is not convertible to a product state by
any gSL transformation.
\end{defn} \noindent

Not all LRE states can be transformed into each other
via gSL transformations.  Thus LRE states can belong to different phases: i.e.\
the LRE states that are not connected by gSL transformations belong to different
phases.  When we restrict ourselves to LRE gapped quantum liquids, those different
phases are nothing but the topologically ordered
phases~\cite{Wtop,WNtop,Wrig,KW9327}.
\begin{defn} \textbf{Topologically ordered phases}\\
Topologically ordered phases
are equivalence classes of LRE gapped quantum liquids under the gSL
transformations.
\end{defn} \noindent

We believe the following observation is true, which provides a support
to the above picture and definition of topologically ordered phases.

\begin{obs}
\label{ob:topoent}
The topological entanglement entropy $S^\text{qua}_\text{topo}$
for topological order is stable under small gSL transformations.
Furthermore, $S^\text{qua}_\text{topo}$ is an invariant for any gSL equivalence class
of topological orders.
\end{obs}

The first sentence of Observation~\ref{ob:topoent} is in fact similar as stating 
that topological orders are stable under imaginary time
evolution. We can also see this as a direct consequence of the quantum
error correction condition given by
Eq.~\eqref{eq:qcode}. Consider any small $\lambda$ and a local Hamiltonian $H$,
for any local operator $M$ and small $\lambda$, the equation
\begin{equation}
\bra{\Psi_i}e^{\lambda H}Me^{\lambda H}\ket{\Psi_j}=C_M\delta_{ij}
\end{equation}
remains to be valid (notice that the 
constant $C_M$ may chance, but the independence of $C_M$ on
the subscripts $i,j$ would remain unchanged).
To see this one only needs to write $e^{\lambda H}$ as $1+ \lambda H$.

Similarly, for symmetry-breaking orders, we have
\begin{cor}
The tri-topological entanglement entropy $S^\text{tri}_\text{topo}$
for symmetry-breaking orders is stable 
under small gSL transformations that do not break symmetry. But unstable
under small gSL transformations that breaks the symmetry.
Furthermore, $S^\text{tri}_\text{topo}$ is not an invariant for any gSL equivalence class
of symmetry-breaking orders.
\end{cor}

As an example, in the transverse Ising model, the gSL transformation which
transforms $\ket{GHZ_+(a)}$ of different $a$ breaks the $\mathbb{Z}_2$ symmetry.
However, $\ket{GHZ_+(a)}$ of different $a$ are in the same gSL equivalent
class, yet with different topological entanglement entropy. 

The second sentence of Observation~\ref{ob:topoent} is more subtle,
as the topological entanglement entropy $S^\text{qua}_\text{topo}$ for topological order
is not an invariant of gSL transformations (as a finite probability $p$ as given 
in Eq.~\eqref{eq:prob} may not exist). This is because that unlike gLU transformations, gSL 
transformations can be taken arbitrarily close to a non-invertible transformation. For instance, take the gSL transformation $W_{N_k}$ as given in Eq.~\eqref{eq:SLeg}. If we allow $\gamma$
to be arbitrarily close to zero, then for any wave function $\ket{\mathcal{V}_{N_k}}$, when applying 
$W_{N_k}\ket{\mathcal{V}_{N_k}}$, it is `as if' we are just projecting everything to $\ket{0}^{N_k}$,
which should not protect any topological order in $\ket{\mathcal{V}_{N_k}}$.

On the other hand, the option to choose $\gamma$ arbitrarily small does not mean any quantum state is gSL convertible to a product state. The key point here is the existence of a finite probably $p$ that is independent of system size $N_k$, as given in Definition~\ref{def:gSLc}. For states with GHZ-form of entanglement, we know that we can always find such a finite probability $p$. 

However, for topological ordered states, there does not exist such a finite probability $p$. In fact,
we have $p\rightarrow 0$ when $N_k\rightarrow\infty$, and furthermore the speed of $p$ approaching $0$ may be exponentially fast in terms of the growth of $N_k$. Therefore $S^\text{tri}_\text{topo}$ shall remain invariant within any gSL equivalent class.

The above idea is further supported by the results known for geometrical entanglement of topological ordered states~\cite{orus2014geometric}.
More precisely, let us divide the system to $m$ non-overlapping local parts, as 
illustrated in Fig.~\ref{qc} for one layer. Label each part by $i$ and write the Hilbert
space of the system by $\mathcal{H}=\bigotimes_{i=1}^M\mathcal{H}_i$. Now
for any normalized wave function $\ket{\Psi}\in\mathcal{H}$, the goal
is to determine how far $\ket{\Psi}$ is from a normalized product state 
\begin{equation}
\ket{\Phi}=\otimes_{i=1}^M\ket{\phi_i}
\end{equation}
with $\ket{\phi_i}\in\mathcal{H}_i$.

The geometric entanglement $E_G(\ket{\Psi})$ is then revealed by the maximal overlap
\begin{equation}
\Lambda_{\max}(\ket{\Psi})=\max_{\ket{\Phi}}|\langle\Phi\ket{\Psi}|,
\end{equation}
and is given by
\begin{equation}
E_G(\ket{\Psi})=-\log\Lambda^2_{\max}(\ket{\Psi}).
\end{equation}
Notice that for $\Lambda_{\max}(\ket{\Psi})$, the maximum is also taken for all the partition
of the system into local parts. 

It is shown in \Ref\cite{orus2014geometric} for a topologically ordered state $\ket{\Psi}$, 
$E_G(\ket{\Psi})$ is proportional to the number of qubits in the system. This means that
the probability to project $\ket{\Psi}$ to any product state is exponentially small in terms
of the system size $N_k$. Therefore one shall not expect $\ket{\Psi}$ to be convertible to
any product state with a finite probability $p$.

In contract, the geometric entanglement for states with GHZ-form of entanglement is
a constant independent of the system size $N_k$. And it remains to be the case for the entire symmetric-breaking phase (see. e.g. \Ref\cite{wei2005global}), which indicates that these GHZ-form states are convertible to product states with some finite probability $p$.  

\subsection{Emergence of unitarity}

The example of toric code discussed in 
Sec.~\ref{sec:LRE} indicates that the gSL and gLU shall give the same
equivalent classes for topological orders, if we restrict ourselves in the case of
LRE states. We believe that this holds in general and summarize it as
the following observation.

\begin{obs}
\label{obs:LREgLU}
For the LRE gapped quantum liquids,
topologically ordered wave functions
are equivalence classes of gLU
transformations.
\end{obs} \noindent
This statement is highly non-trivial since, in the above, the concept of LRE
and topologically ordered wave functions are defined via non-uintary gSL
transformations.  Observation~\ref{obs:LREgLU} reflect one aspect of emergence
of unitarity in topologically ordered states.

The locality structure of the total Hilbert space is described by the tensor
product decomposition: $\mathcal{H}=\otimes_{i=1}^m \mathcal{H}_i$ where
$\mathcal{H}_i$ is the local Hilbert space on $i^\text{th}$ site.  The inner
product on $\mathcal{H}$ is compatible with the locality structure if it is
induced from the inner product on $\mathcal{H}_i$.  A small deformation of the
inner product on $\mathcal{H}_i$ can be induced by a small gSL transformation.

Consider an orthonormal basis $\{\ket{\Psi}_i\}$ 
of a topologically ordered 
degenerate ground-state subspace, where $\langle\Psi_i|\Psi_j\rangle=\delta_{ij}$.
Since the a small gSL transformation does not change the orthonormal property 
$\langle\Psi_i|\Psi_j\rangle=\delta_{ij}$, thus a small deformation of
the inner product also does not change this orthonormal property.
This is another way of stating that small gSL transformations
can be realized as gLU transformations for topologically ordered degenerate
ground states, which represent another aspect of emergence of unitarity.

We can then summarize the above argument as 
the following observation.
\begin{obs}
For an orthonormal basis $\{\ket{\Psi_i}\}$ 
of a topologically ordered 
degenerate ground-state subspace, 
the orthonormal property 
$\langle\Psi_i|\Psi_j\rangle=\delta_{ij}$ for $i\neq j$ is invariant under a small deformation of the
inner product, as long as the inner product is compatible with the locality
structure of the total Hilbert space. 
\end{obs} 
\noindent That is, for a given orthonormal basis $\{\ket{\Psi_i}\}$,
$\langle\Psi_i|\Psi_j\rangle$ does not change, up to an overall factor, under a small deformation of the inner
product.

We can also view the emergence of unitarity from the viewpoint of
imaginary-time evolution.  In particular, if one imaginary-time evolution leads
to degenerate ground states for a topologically ordered phase, a slightly
different imaginary-time evolution will lead to another set of degenerate
ground states for the same topologically ordered phase.  The two sets of
degenerate ground states are related by a unitary transformation.  In  this
sense, topological order realizes the emergence of unitarity at low energies.

\section{Summary and discussion}

In this work we have introduced the concept of gapped quantum liquids, which is
a special kind of gapped quantum states.  There exist gapped quantum states
which are not gapped quantum liquids, such as 3D gapped states formed by stacking
2D $\nu=1/3$ fractional quantum Hall states.  The cubic code may provide such
an example.  We show that topologically ordered states, whose Hamiltonians have
stable ground state degeneracy against any local perturbations, belong to
gapped quantum liquids. On-site-ymmetry-breaking states are also gapped quantum
liquids, whose Hamiltonians have unstable ground-state degeneracy.  This result
implies that it is incorrect to regard every gapped state without symmetry as a
topologically ordered state. There are more exotic gapped states than
topologically ordered states.  This result also allows us to give a more
precise definition of topological order.

We have shown that gLU classes for stable gapped quantum liquids have a
one-to-one correspondence to topological orders. For unstable gapped quantum
liquids, gLU transformations assign symmetry-breaking orders to non-trivial
classes.  We have introduced a new topological entanglement entropy
$S^\text{tri}_\text{topo}$ that can detect symmetry-breaking orders.  As
topological entanglement entropies $S^\text{tri}_\text{topo}$ and
$S^\text{qua}_\text{topo}$ are invariants under gLU transformations, we can use
them to study both topological orders and symmetry-breaking orders.

We introduce the idea of gSL transformations and define gSL convertibility of
quantum states.  This convertibility is a partial order (in terms of set
theory) on quantum states and it connects symmetry-breaking ground states to
the product states. In this sense we re-define the concept of short/long range
entanglement and have shown that only topologically ordered states are
long-range entangled, in a sense that they are not convertible to product
states by gSL transformations. 

We show that the topological entanglement entropies $S^\text{tri}_\text{topo}$
and $S^\text{qua}_\text{topo}$ for topological order are stable under gSL
transformations, and are invariants within any gSL equivalent class, although it
is not an invariant of gSL transformations in general (which may be arbitrarily
close to a projection onto a product state).  On the other hand,
$S^\text{tri}_\text{topo}$ is not stable for symmetry-breaking orders. We
further show that for the LRE gapped quantum liquids (i.e.\ topological orders),
gSL equivalent classes coincide with the gLU equivalent classes.  This is
consistent with the observation that gLU classes for stable gapped quantum
liquids have a one-to-one correspondence to topological orders, which realizes
the emergence of unitarity at low energies.
This result reveals to true essence of topological order and long-range entanglement: the emergence of unitarity at low energies. 

\section*{Acknowledgements}
We thank Zhengfeng Ji and Nengkun Yu for helpful discussions. BZ is supported
by NSERC.  XGW is supported by NSF Grant No.  DMR-1005541 and NSFC 11274192.
He is also supported by the BMO Financial Group and the John Templeton
Foundation.  Research at Perimeter Institute is supported by the Government of
Canada through Industry Canada and by the Province of Ontario through the
Ministry of Research.

\end{document}